\documentclass{mn2e}
\usepackage[utf8x]{inputenc}
\usepackage[english]{babel}
\usepackage{times}
\usepackage[T1]{fontenc}
\usepackage{array}
\usepackage{amsmath}
\usepackage{graphicx}

\newcommand{\w}{\bf}    

\title[Magnetic field in molecular clouds]{Magnetic field generation in galactic molecular clouds}
\author[Ya.~N. Istomin, A.~Kiselev]{Ya.~N. Istomin $^{1,2}$, A.~Kiselev $^{1,3}$ \\
${1}$ P.N.~Lebedev Physical Institute, Leninsky Prospect 53, Moscow 119991, Russia \\
${2}$ E-mail: istomin@lpi.ru \\
${3}$ E-mail: kiselevalexs@gmail.com
}

\begin{document}
\date{}
\pagerange{\pageref{firstpage}--\pageref{lastpage}} \pubyear{2013}
\maketitle
\label{firstpage}

\begin{abstract}
We investigate the magnetic field which is generated by turbulent motions
of a weakly ionized gas. Galactic molecular clouds give us an
example of such a medium. As in the Kazantsev-Kraichnan model we
assume a medium to be homogeneous and a neutral gas velocity field
to be isotropic and delta-correlated in time. We take into
consideration the presence of a mean magnetic field, which defines
a preferred direction in space and eliminates  isotropy of
magnetic field correlators. Evolution equations for the
anisotropic correlation function are derived. Isotropic cases with
zero mean magnetic field as well as with small mean magnetic field
are investigated. It is shown that stationary bounded solutions
exist only in the presence of the mean magnetic field for the
Kolmogorov neutral gas turbulence. The dependence of the magnetic 
field fluctuations amplitude on the mean field is
calculated. The stationary anisotropic solution for the magnetic
turbulence is also obtained for large values of the mean magnetic
field.
\end{abstract}


\section{Introduction}
The standard theory of cosmic rays (CR) formation suggests that
primary CR consist mainly of protons and contain no antimatter.
During their propagation in the Galaxy primary CR interact with
protons of galactic gas, resulting in production of secondary CR,
including antiprotons and positrons. The secondary particles
energy spectrum, calculated in the framework of this theory, falls
down with energy by a power law manner. The antiparticles to
particles ratio should behave in the same way (Moscalenko \&
Strong 1998). However, recent antimatter observations by 
\textit{PAMELA} satellite detected an excess of positrons with energies $10-100$ 
GeV (Adriani et al. 2009). In this range, the ratio $e^{+}
/(e^{-} + e^{+})$ is about $ 10\%$ and increases with energy.
These observations attracted much attention. Several theoretical
explanations for this effect were proposed, among them the positrons
generation in pulsars and in the annihilation process of dark matter
particles. Another mechanism for the positrons generation in the
Galaxy is also possible, that is acceleration of charged particles
in giant molecular clouds and secondary CR production there. This
mechanism was discussed by Dogiel et al. (1987, 2005), long before
the launch of the \textit{PAMELA} satellite in 2006. 

The particle acceleration in molecular clouds takes place due to turbulent motion of a partially ionized gas inside them. 
This motion switches on the dynamo mechanism of a magnetic field generation.
Besides the magnetic field $\w B$ the electric field $\w E$ also appears, 
$E = - ({\w u} - \nu_m \nabla) \times {\w B} /c$. Here ${\w u}$ is plasma velocity and
$\nu_m$ is the magnetic viscosity.
Moving in this stochastic electric field, protons and electrons gain energy and can be accelerated up to 
energies~$> 10$ GeV (Dogiel et al. 1987). In the presence of magnetic field one can describe motion of relativistic charged particles as diffusion in coordinate and momentum spaces. To find diffusion coefficient and, after that, maximum energy and spectrum of accelerated particles one need to know properties of the magnetic field, for example, its pair correlators (see Shalchi 2009; Dogiel et al. 1987), which will be calculated below.
Dogiel et al. (2005) predicted the positron excess in GeV energy range.
Appearance of appropriate observations requires a detailed
investigation of particle acceleration in molecular clouds, with taking into account modern
data.

\begin{figure}
\includegraphics[width=9cm]{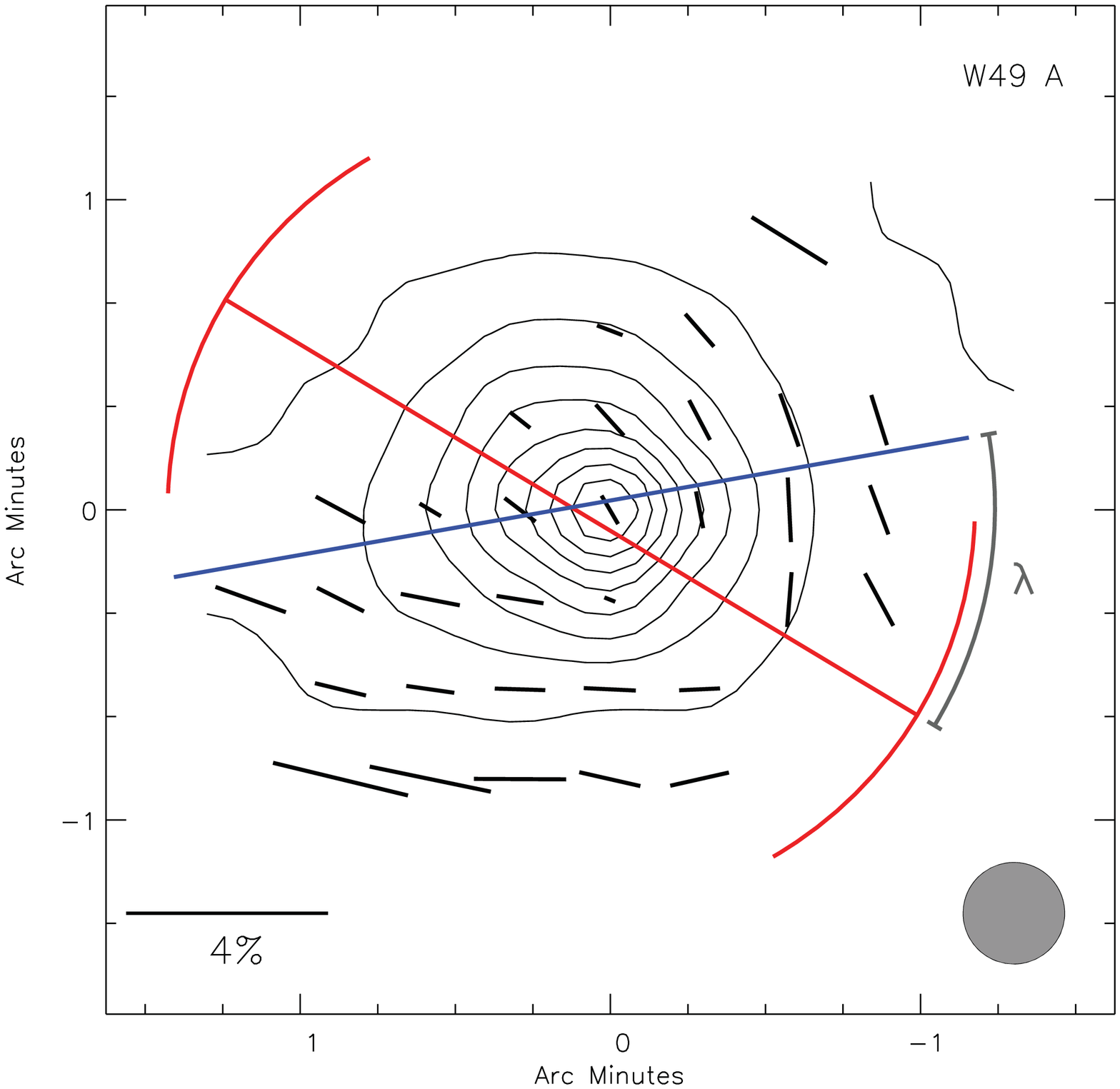}
\caption{Observations of molecular cloud W49A. Contours represent the levels of 
the radio flux. Small black line-segments shows the
direction of the magnetic field at the point. Figure is taken from Tassis et al. (2009).}
\label{F:pol}
\end{figure}

Molecular clouds are clusters of molecular hydrogen with a complex
inhomogeneous structure (Larson 2003). Their dimensions may reach
$100$~parsecs, masses are up to $10^6 \, M_\odot$. Gas concentration
in molecular clouds $N_n$ is about 
$(10^{2} - 10^{3}) \, \text{cm}^{-3}$, the gas temperature is $T = (10-50) \, \text{K}$.
According to observations, gas is strongly turbulent. The
turbulence has a power law Kolmogorov-like spectrum. In addition,
a gas is partially ionized $N_i/N_n=10^{-8}-10^{-5}$. In such a
system stochastic magnetic field arises, as will be shown below.
The only way to measure directly the magnetic field strength is
the Zeeman effect. Zeeman observations were carried out recently
for many clouds, their results are summarized in the paper by
Crutcher (2010). Typical values of the magnetic field projection on
the line of sight are $10-30 \, \mu G$ for the molecular clouds
cores. Polarization observations (Tassis et al. 2009), carried
out for several molecular clouds, showed that magnetic field
directions in distant points of a cloud may be similar, 
Fig.~\ref{F:pol}. It appears that a mean homogeneous magnetic field
exists in clouds together with a stochastic field, produced by the
turbulence. In papers (Dogiel et al. 1987, 2005) a mean magnetic
field was assumed to be zero. In present paper we treat the
problem of magnetic field generation in weakly ionized turbulent
gas with a nonzero mean magnetic field. No assumptions about the
ratio of a mean to a fluctuating fields are made.

This paper consists of five parts. In Section 2 we write equations describing a gas and a magnetic field in molecular clouds.
Evolution equations for magnetic field pair correlators are derived in Section 3.
Their stationary solutions are obtained in Section 4. Three cases are considered in detail: a) zero mean magnetic field, b) small mean field
and c) large mean field.
In Section 5 we compare our results with that obtained by other authors.
Summary is compiled in Section 6.

\section{Magnetohydrodynamics equations for weakly ionized gas}
For description of gas motion in molecular clouds one can use
two-fluid hydrodynamic equations. We denote  velocities of a neutral
and an ionized gas components as $\w v$ and $\w u$
respectively. Magnetic viscosity $\nu_m$ in molecular clouds is
much less than the kinematic viscosity $\nu$ (see Dogiel et al. 1987).

We consider gas motions on the scales $L$ corresponding to the
inertial range $L_\nu<L<L_0$, where $L_0$ is determined by the
size of the system, and $L_\nu$ corresponds to the viscous scale.
Typical values for molecular clouds are $L_0 = 10^{19}$ cm,
Reynolds number is $Re=10^8$, hence for the Kolmogorov turbulence
$L_\nu = L_0 Re^{-3/4} = 10^{13}$ cm. In this range of scales the
viscosity can be neglected. Turbulent velocity at small scales is
less than the sound velocity, it reaches the value of sound
velocity only at large scales. So we assume gas to be
incompressible, because subsonic gas motions can be considered to
be incompressible. Then equations for the ionized component motion
and the magnetic field are
\begin{eqnarray} \label{E:gidro}
&&\frac{\partial \w u}{\partial t} + {(\w u \nabla)u} = \frac{1}{\rho_i}\left[-{\w \nabla} P_i +
\frac{\w{(\nabla \times B) \times B}}{4 \pi}\right] - \\ \nonumber
&& - \mu_{in} ({\w u-v}),  \\  \label{E:induction_eq}
&&\frac{\partial \w B}{\partial t} = \w \nabla \times (u \times B),   \\ \nonumber
&&\nabla \w u = 0,   \\ \nonumber
&&\nabla \w B = 0,
\end{eqnarray}
where $P_i$ and $\rho_i$ are the pressure and the density of
ionized component, $\mu_{in}$ is ion-neutral collision rate. In
contrast to the usual magnetohydrodynamics, in this case an
external force in the form of friction between the ionized and the
neutral gas components is present. Indeed, estimations give
\begin{equation}
\mu_{ni} = 10^{-13} \,\text{s}^{-1}, \qquad
\mu_{in} = 2 \cdot 10^{-6} \,\text{s}^{-1} ,
\end{equation}
whereas the turbulent fluctuation frequencies lie in the range
\begin{equation}
\omega_{min} = 10^{-12} \,\text{s}^{-1}, \qquad
\omega_{max} = 10^{-8} \,\text{s}^{-1}.
\end{equation}
This implies an important condition
\begin{equation}
\mu_{ni} \ll \omega \ll \mu_{in}.
\end{equation}
This means that the neutral gas does not feel the presence of an ionized component. Ions motion, by contrast, is completely determined by the
motion of a neutral gas. Numerical estimates of the characteristic frequencies of the problem were discussed in detail by Dogiel et al. (1987).

Thus we can treat the motion of the neutral component to be known, it coincides with the ordinary hydrodynamic turbulence.
For an ionized component only two forces are essential: the force of friction on a neutral gas and the Lorentz force, caused by a magnetic field.
As we will see below, the pressure $ P_i $ can be neglected in comparison with the pressure of the magnetic field.
Therefore the motion equation for an ionized component becomes
\begin{equation}
\frac{\w (\nabla \times B) \times B}{4 \pi \rho_i} - \mu_{in} \w (u-v) =0.
\end{equation}
Let us denote
\begin{equation}
a = \frac{1}{4\pi \rho_i \mu_{in}}.
\end{equation}
As far as we consider a gas to be incompressible, $a =$ const. Expressing the  velocity of an ionized component in terms of velocity of neutrals
and substituting it to the induction equation (\ref{E:induction_eq}), we obtain
\begin{equation} \label{E:dBdt}
\frac{\partial \w B}{\partial t} = {\w \nabla \times(v \times B)}
-  a \w \nabla \times (B \times (\nabla \times B) \times B) .
\end{equation}
This equation gives the dependence of the magnetic field on the velocity $\w v$ of a neutral gas.

\section{Derivation of evolution equation}
\subsection{Tensor structure of correlators}
To describe the turbulent motion of a neutral gas and obtain closed equations for magnetic field correlators, we use solvable model proposed by Kazantsev (1968) and Kraichnan (1968). We assume neutral gas velocity ${\w v}(t)$ to be a Gaussian stochastic process with zero mean value,
$\langle {\w v} \rangle =0$. All information about it is contained in the pair correlation function
$\langle v_i({\w x},t) v_j({\w x + r},t') \rangle$.
The angle brackets here and below denote averaging over an ensemble of realizations.
We assume a neutral gas to be a homogeneous isotropic medium, so the pair correlation function can depend only on $\w{r}$.
We consider the velocity field to be delta-correlated in time,
\begin{equation}\label{E:delta_corr}
\langle v_i({\w x},t) v_j({\w x + r},t') \rangle =v_{ij}({\w r}) \tau_c \delta(t-t'),
\end{equation}
and mirror-symmetric. It possesses no helicity, and its correlation tensor is symmetric with respect to the interchange of indices
$v_{ij}({\w r}) = v_{ji}({\w r})$. In this case one can construct the tensor structure of the correlation function from only two second-rank tensors
\begin{equation}
v_{ij} =2 V(r) \delta_{ij} + S(r) ( \delta_{ij} - \frac{r_i r_j}{r^2}).
\end{equation}
The factor $2$ in the first term of the right hand side is written for convenience. From the incompressibility condition
$\nabla \w{v} =0$, i.e. $\partial_i v_{ij} = \partial_j v_{ij} =0$, we obtain
\begin{equation}\label{E:compessibility}
S = r V',
\end{equation}
where prime denotes the derivative with respect to $r$, $V'=\partial V / \partial r$. This relation leads us to the general form of the
correlation tensor
\begin{equation}\label{tenzor_v}
v_{ij}(r) = 2 V(r) \delta_{ij} + r V'(r) \left(\delta_{ij} - \frac{r_i r_j}{r^2}\right).
\end{equation}
Thus, the neutral gas turbulence is described by one scalar function $V(r)$.
A common method to handle this problem is to pass to the Fourier space
\begin{equation}
{\w v(r},t) = \int {\w v(k},t) \exp(i{\w kr})d{\w k}.
\end{equation}
Indeed, the correlation tensor structure becomes simplier
\begin{equation}
\langle v_i({\w k},t)v_j({\w k'},t') \rangle =\overline{V}(k) \tau_c \delta(t-t')\left(\delta_{ij} - \frac{k_i k_j}{k^2}\right)\delta({\w k+k'}) .
\end{equation}
The factor $\delta({\w k+k'})$ arises from homogeneity, and the tensor structure is uniquely determined by isotropy and incompressibility condition
$k_i v_i(k)=0$.
Functions $V(r)$ and $\overline{V}(r)$ are related by
\begin{equation}\label{E:connection}
\overline{V}(r) = 3 V(r) + r V'(r),
\end{equation}
and $\overline{V}(k)$ is a Fourier transform of $\overline{V}(r)$,
\begin{equation}
\overline{V}(r) = \int \overline{V}(k) \exp(i ({\w kr})) d \w k.
\end{equation}
Now let us consider correlators of the magnetic field. We assume magnetic field to be a Gaussian stochastic process too.
But it has nonzero mean value. Let us denote its mean and fluctuating components by $\w H$ and $\w b$ respectively:
\begin{equation}
\w B= H+ b , \qquad \langle B \rangle= H.
\end{equation}
We suppose mean magnetic field to be constant in space ${\w H} = \text{const}({\w r})$.
One can look for the evolution equation for ${\w H}(t)$, using the technique described below, and get $\partial{\w H} / \partial t=0$. Hence the
value of the mean field ${\w H} = \text{const}({\w r}, t)$ is an external parameter of the problem. 

Strictly speaking, magnetic field $\w B$, generated by the Gaussian stochastic field $\w v$, is not pure Gaussian. Its properties are not completely described by the second order correlation function. But significant difference appears in higher than second order moments, and for investigation of second order moment evolution one can assume $\w B$ to be Gaussian, see, for example, the paper by Brandenburg \& Subramanian (2000). 

To describe fluctuating component of
the magnetic field, we introduce its pair correlator
$ \langle b_i({\w x},t)b_j({\w x+r},t) \rangle $,
which is similar to the velocity correlator, but the average is taken at the same time moments. Our aim is to establish the evolution equation for
this correlator. Since
\begin{equation}
\frac{\partial}{\partial t} \langle b_i b_j \rangle =
\langle \frac{\partial b_i}{\partial t}b_j \rangle +
\langle b_i  \frac{\partial b_j}{\partial t} \rangle,
\end{equation}
one have to calculate $\partial {\w b}/\partial t$. We suppose that
$\langle b_i b_j b_k \rangle=0$. Averaging Eq.~(\ref{E:dBdt}), subtracting the resulting equation from Eq.~(\ref{E:dBdt}), we get
\begin{eqnarray} \label{dbdt_r}
&&\frac{\partial \w b}{\partial t} = {\w \nabla \times [v \times H] + \nabla \times [v \times b] }
-a \w \nabla \times \Bigl(  H \times[j \times H] + \nonumber\\
&& + \w H \times[j \times b] +  b \times[j \times H] +  b \times[j \times b] \Bigr) - \nonumber\\
&& -{\w \nabla \times \langle v \times b \rangle} + a \w \nabla \times \langle H \times[j \times b] +  b \times[j \times H] \rangle,
\end{eqnarray}
where we use the notation $\w j=\nabla \times b$ for short.

Last three terms (last line) in Eq.~(\ref{dbdt_r}) give no contribution to
$\partial \langle b_i b_j \rangle / \partial t$, since
$ \langle \langle \ldots \rangle b_i \rangle =
\langle \ldots \rangle \langle b_i \rangle =0,  $
so they can be ignored. Similarly we omit terms in Eq.~(\ref{dbdt_r}), which are proportional to $b^2$, because their contribution to the derivative
$\partial \langle b_i b_j \rangle / \partial t$ is
proportional to $\langle b^3 \rangle=0$.

\subsection{The case with zero mean field}\label{SS:deriv_iso}
To begin with we consider the case when the mean field is absent, $H =0$. Then all correlators are isotropic.
Maxwell equation $\nabla {\w b}=0$ is similar to the incompressibility equation $\nabla {\w v}=0$, so tensor structure of the magnetic field
correlator for the non-helical case is similar to the velocity correlator (\ref{tenzor_v})
\begin{equation}\label{CorrSymB}
\langle b_i({\w x},t)b_j({\w x+r},t) \rangle = 2 Q(r) \delta_{ij} + r Q'(r)\left(\delta_{ij} - \frac{r_i r_j}{r^2}\right).
\end{equation}
In the Fourier space the correlator is
\begin{equation}
\langle b_i({\w k},t)b_j({\w k'},t) \rangle =\overline{Q}({\w k})\left(\delta_{ij} - \frac{k_i k_j}{k^2}\right) \delta(\w k+k'),
\end{equation}
where the relation between $Q(r)$ and $\overline{Q}(k)$ is similar to (\ref{E:connection}). Due to the isotropy
$\overline{Q}({\w k}) = \overline{Q}(k)$.
To get $\partial \overline{Q}(k) / \partial t$ we apply Fourier transform to Eq.~(\ref{dbdt_r}),
drop terms which do not contribute to the value of
$\partial \langle b_i b_j \rangle / \partial t$, put $H=0$ and obtain
\begin{equation}  \label{dbdt_k}
\begin{split}
\frac{\partial \w b(k)}{\partial t} &= i \int d{\w q \, k \times [v(q) \times b(p)] } + \\
 & + a \int d{\w k_1} d {\w k_2} \, {\w k \times \{b(k_1) \times ([k_2 \times b(k_2)] \times b(k_3) )} \}.
\end{split}
\end{equation}
In the first term of the right hand side $\w p+q=k$, in the second term
$\w k_1+k_2+k_3 = k$.
One can see the following structure of correlator's derivative
\begin{equation}\label{struct}
\frac{\partial}{\partial t} \langle b b \rangle \simeq
 \langle vb^2 \rangle + \langle b^4 \rangle,
\end{equation}
where only the magnetic field $b$ and the velocity $v$ are shown, and all tensor indices are dropped.
We assume random processes $\w v$ and $\w b$ to be Gaussian. So, to split the correlators in the first term of Eq.~(\ref{struct}) one should
use the Furutsu-Novikov formula (see Furutsu 1963; Novikov 1965; Klyatskin 2005). It states that if some functional $R[{\w v}]$ depends on the random process ${\w v}({\w k},t)$
as a solution of some differential equation, one can split correlator
\begin{equation}
\langle v_i({\w k},t) R[{\w v}] \rangle= \int d{\w k'} dt' \, \langle v_i({\w k},t) v_j({\w k'},t') \rangle
\left< \frac{\delta R[{\w v}]} {\delta v_j({\w k'},t')} \right>,
\end{equation}
where $\langle \delta R[{\w v}] / \delta v_j({\w k'},t') \rangle$ is a functional derivative. To compute it one should use the equation
which specifies the dependence of $R$ on ${\w v}$. In our case it is the evolution equation (\ref{dbdt_k}). Since the random process ${\w v}$ is
assumed to be delta-correlated in time, and at $t'>t$ causality principle states that
$ \delta b_i({\w k},t) / \delta v_j({\w k'},t') = 0$, we need to know only the value of
\begin{equation}  \label{dbdtFunctional_k}
\lim_{t' \to t-0} \quad \frac{\delta b_i({\w k},t)} {\delta v_j({\w k'},t')}  = i k_m  (\delta_{ij}b_m^{{\w k-k'}} - \delta_{jm}b_i^{\w k-k'} ).
\end{equation}
In the right hand side the argument is written as a superscript for clarity. In the second term of Eq.~(\ref{struct}) one can presents forth order
correlator as a product of pair correlators
\begin{equation}
\langle b_1 b_2 b_3 b_4 \rangle = \langle b_1 b_2 \rangle \langle b_3 b_4 \rangle + \langle b_1 b_3 \rangle \langle b_2 b_4 \rangle +
\langle b_1 b_4 \rangle \langle b_2 b_3 \rangle.
\end{equation}
Let us write out the intermediate result obtained after splitting of correlators. To do this, we use the notations
\begin{equation}
\Omega_{ij}^{\w k} = \delta_{ij} - \frac{k_i k_j}{k^2},
\end{equation}
and ${\w p}=\w k-q$. We obtain
\begin{eqnarray}
&&\frac{\partial \overline{Q}(k)}{\partial t} \Omega_{ij}^{\w k} =
 2 a \int d{\w q} \, \overline{Q}(k) \overline{Q}(q)
\Bigl(  -k^2 \Omega^{\w q}_{is} \Omega^{\w k}_{sj} - \\ \nonumber &&
-k_m k_s \Omega^{\w q}_{ms} \Omega^{\w k}_{ij} +
k_m k_i \Omega^{\w q}_{ms} \Omega^{\w k}_{sj} \Bigr)
- \tau_c k_m \int d{\w q} \, \overline V (q) \overline{Q}(k) \times \\ \nonumber &&
\times \Bigl( \Omega^{\w q}_{is} p_n (\delta_{ms} \Omega_{nj}^{\w k} - \delta_{ns} \Omega_{mj}^{\w k}) - (i \leftrightarrow m)\Bigr)
 + \tau_c k_m \times \\ \nonumber &&
\times \int d{\w q} \, \overline V (q) \overline{Q}(p) \,
\Bigl( \Omega^{\w q}_{is} k_n (\delta_{js} \Omega_{nm}^{\w p} - \delta_{ns} \Omega_{mj}^{\w p}) - (i \leftrightarrow m) \Bigr).
\end{eqnarray}
Tensors in right hand side of last equation contain vectors $q_i$
and $p_i$, which are absent in the left hand side. To extract the
factor $\Omega_{ij}^{\w k}$ in the right hand side one can use
isotropy of  $\overline V (q)$ and $\overline{Q}(k)$, then average
first two integrals over the sphere $|\w q| = \text{const}$ and
the third integral over a circle, determined by conditions $ |\w
q| = \text{const}$ and $ |\w p| =\text{const}$. After that tensor
factors are cancelled and we get scalar evolution equation
\begin{equation}\label{E:iso_k}
\begin{split}
\frac{1}{\tau_c}\frac{\partial \overline Q(k)}{\partial t} =&
- \frac{2}{3} k^2 \overline Q(k)  \int \, \overline V(q) d{\w q} + \\
&+ k^2 \int \, \overline V(q) \overline Q(p) \left[1 -\frac{({\w kp})({\w kq})({\w pq})}{k^2 p^2 q^2}\right]d{\w q} - \\
&- \frac{2}{3\pi \rho_i \mu_{in} \tau_c} k^2 \overline Q(k)  \int \,
\overline Q(q) d{\w q},
\end{split}
\end{equation}
where ${\w k}=\w p+q$. This equation was derived firstly by Dogiel et al. (2005).
Let's introduce the notation
\begin{equation}
\lambda=\frac{1}{3\pi \rho_i \mu_{in} \tau_c} \int \, \overline{Q}(q) d{\w q}  .
\end{equation}
The value of $\lambda$ is proportional to
$b_0^2 = \langle b_i({\w x})b_i({\w x}) \rangle$  which is the energy of the fluctuating magnetic field.

Since solving the integral equation on $\overline{Q}(k)$ in the $k$-space is rather complicated problem, let us turn back to the $r$-space
and obtain the differential equation for $Q(r)$. To do this, we apply inverse Fourier transform to the Eq.~(\ref{E:iso_k}).
One can express the value of parameter $\lambda$ in terms of $Q(r)$
\begin{equation}
\lambda  = \frac{4}{3} \frac{a}{\tau_c} \overline{Q}(r=0) =
\frac{2 a}{\tau_c} (2 Q(r)+\frac{2}{3} r Q'(r))|_{r \to 0}  .
\end{equation}
Also note that
\begin{equation}
V(0)  = \frac{1}{3} \int \, \overline V(q) d \w q  .
\end{equation}
We proceed from functions $\overline{Q}(r)$, $\overline V(r)$ to $Q(r)$, $V(r)$ according to Eq.~(\ref{E:connection}), use the spherical symmetry
of the correlation functions and get after some algebra
\begin{equation}\label{E:iso_r}
\begin{split}
\frac{1}{2\tau_c} \frac{\partial Q(r)}{\partial t} =&
\left(V(0) - V(r) + \lambda \right) (Q'' + \frac{4Q'}{r}) - \\
&- V' Q' - \frac{1}{r}(4V' + r V'')Q .
\end{split}
\end{equation}
One of the key points of this derivation is averaging over a sphere (or circle), when we use the isotropy of the functions $\overline V(q)$ and $\overline{Q}(k)$.
In the anisotropic case, in the presence of the mean field, for example, this averaging fails. Thus, we conclude that in general case
it is necessary to derive and solve the evolution equation in $r$-space. It is no need to apply a Fourier transform. Tensor structure of
correlators (\ref{tenzor_v}) in this method is more complicated, but there arise no integral equations and additional vectors (as $p_i$, $q_i$) in
the tensor structure.

Indeed, the equation (\ref{E:iso_r}) can be obtained directly in
the $r$-space, by analogy with the above derivation. We will
discuss its solution in the section \ref{SS:sol_iso}.

\subsection{The general case of a non-zero mean field $H \ne 0$}
In this part we consider general case when the mean magnetic field
is present. The correlation function of magnetic fluctuations
becomes anisotropic. We assume magnetic field to have no helicity,
so its correlation tensor $\langle b_i b_j \rangle$ is symmetric
with respect to the interchange of indices. We suppose that it has the
form
\begin{equation}\label{corr_H}
\begin{split}
\langle b_i({\w x},t)b_j({\w x+r},t) \rangle &= A({\w r}) \delta_{ij} + B({\w r}) n_{i}n_j + \\
&+ C({\w r}) (n_i h_j+n_j h_i) + D({\w r}) h_i h_j ,
\end{split}
\end{equation}
where ${\w n}={\w r}/r$, ${\w h}={\w H}/H$ are unit vectors in the directions of $\w r$ and $\w H$ respectively.

The most general form of the correlation tensor of the second order in the presence of a preferred direction of the mean magnetic field
is given in the paper by Matthaeus \& Smith (1981). However, we restrict our attention to terms, specified in Eq.~(\ref{corr_H}). As we will
see below this assumption is consistent, in the right hand side of evolution equations only the same tensor terms arise.

All scalar functions are no longer spherically symmetric, but they are axially symmetric with respect to the  direction of $\w H$.
We choose this direction as $z$-axis.
Below spherical coordinates $(r, \, \mu=\cos \theta)$ are used, where $\theta$ is the polar angle between
$\w n$ and $\w h$.

From the symmetry of the correlator and homogeneity of space, functions $A, B, D$ should be symmetric with respect to~$\mu$,
$A(r,\, \mu) = A(r,\, -\mu)$, whereas function $C$ should be anti-symmetric
$C(r,\, \mu) = -C(r,\, -\mu)$.

Condition $\nabla {\w b}=0$ gives us two relations
\begin{eqnarray}\label{E:connections}
&& A'_r - \frac{\mu}{r} A'_\mu + B'_r +\mu C'_r + \frac{1-\mu^2}{r}C'_\mu + \frac{2 B}{r} -
\frac{\mu C}{r}=0 \nonumber \\
&& \frac{1}{r}A'_\mu + C'_r + \mu D'_r + \frac{1-\mu^2}{r}D'_\mu + \frac{3C}{r} =0.
\end{eqnarray}
The derivation of the evolution equations is similar to that performed in the previous section, but now we work in the $r$-space. We use
Eq.~(\ref{dbdt_r}) instead of Eq.~(\ref{dbdt_k}) and the functional derivative
\begin{equation}\label{dbdtFunctional_r}
\frac{\delta b_i({\w r},t)} {\delta v_j({\w r'},t-0)}  = \frac{\partial}{\partial r_m}
\left( \delta({\w r-r'}) (B_m({\w r})  \delta_{ij} - B_i({\w r}) \delta_{jm} )  \right),
\end{equation}
where $\w B=H+b$ is total magnetic field, instead of the formula (\ref{dbdtFunctional_k}).
As before, we use the relation
\begin{equation}
\langle b_i({\w r}) \frac{d}{dr} b_j({\w r}) \rangle = 0.
\end{equation}
which is due to the symmetry of the correlator and homogeneity of space.
Finally one principally can write out the bulky system of evolution equations (for the functions
$A, B, C, D$), which contains the second order derivatives in the right hand side. It must be solved taking into account relations
(\ref{E:connections}). To solve such a system would be very difficult.
To simplify this system, one can use
the structure of the Eq.~(\ref{E:dBdt}), and split the system of equations with second-order spatial derivatives
into two systems with first-order derivatives.
Namely, we replace one of the unknown functions by 
\begin{equation}
\tilde D =D + H^2,
\end{equation}
introduce the notations
\begin{eqnarray}\label{E:def_XY}
X &=& a (D(0)+H^2) = a \tilde D(0) \\ \nonumber
Y(r) &=& -2 a A(0)+ \tau_c( V(r) - V(0) +\frac{1}{2}rV')
\end{eqnarray}
and get
\begin{equation}\label{E:struct_tens_evol}
\begin{split}
&\frac{\partial}{\partial t} \langle b_k({\w x})b_j({\w x+r)} \rangle =
e_{ksi} \partial_s \Bigl( f_1 h_m e_{ijm} + f_2 [{\w nh}]_i h_j + \\
& f_3 [{\w nh}]_i n_j + f_4 n_m e_{ijm} \Bigr)
  + \Bigl( j \leftrightarrow k, {\w r \leftrightarrow (-r)} \Bigr) = 0,
\end{split}
\end{equation}
where $e_{ksi}$ is the completely anti-symmetric pseudo-tensor, and
\begin{eqnarray}\label{E:def_f_i}
f_1&=&
X (\mu A'_r + \frac{1-\mu^2}{r} A'_\mu - \frac{\mu B+C}{r}) +Y (-\frac{1}{r}A'_\mu+\frac{C}{r}) - \nonumber \\  &&
 - (3V'+r V'') (C+\mu \tilde D)\\[0.2cm] \nonumber
f_2&=&
-X (A'_r -\frac{\mu}{r} A'_\mu - \frac{B}{r}) +(X-Y)\frac{1}{r}C'_\mu - \\ \nonumber &&
 - (X-Y+r V')\tilde D'_r + (X-Y)\frac{\mu}{r} \tilde D'_\mu + V' \tilde D \\[0.2cm] \nonumber
f_3&=&
(X-Y)\frac{1}{r}B'_\mu - (X-Y+r V')C'_r +
(X-Y)\frac{\mu}{r}C'_\mu + \\ \nonumber &&
+ \frac{1}{r}(X-Y+r^2 V'')C - (V'-r V'') \mu \tilde D  \\[0.2cm] \nonumber
f_4&=&
(r V'-Y) A'_r + Y(\frac{\mu}{r}A'_\mu+ \frac{B}{r}) - V' A - \\ \nonumber &&
- (3V'+r V'')(A+B+\mu C).
\end{eqnarray}
Let us note that the correlator of the form (\ref{corr_H}) with function $\tilde D$ instead of $D$ in the right hand side describes the correlator of the
total magnetic field $\langle B_i B_j \rangle$.
The system of evolution equations with nonzero mean field is not homogeneous, it contains a "source" term, which is
proportional to $H^2$. However, the system with function $\tilde D$ is homogeneous, but the
term with $H^2$ arises in boundary conditions.

\noindent
From Eq.~(\ref{E:struct_tens_evol}) one can obtain evolution equations:
\begin{eqnarray}\label{E:evolution}
\frac{1}{2} \frac{\partial A}{\partial t} &=&
\mu \partial_r f_1+ \frac{1-\mu^2}{r} \partial_\mu f_1 + \partial_r f_4 + \frac{f_4}{r} \\[0.2cm] \nonumber
\frac{1}{2} \frac{\partial B}{\partial t} &=&
\mu \partial_r f_3+ \frac{1-\mu^2}{r} \partial_t f_3 - 2\mu\frac{f_3}{r} - \partial_r f_4 + \frac{\mu}{r} \partial_\mu f_4 + \frac{f_4}{r} \\[0.2cm] \nonumber
\frac{1}{2} \frac{\partial C}{\partial t} &=&
\frac{1}{2} \Bigl( -\partial_r f_1 + \frac{\mu}{r} \partial_\mu f_1 + \mu \partial_r f_2 + \frac{1-\mu^2}{r} \partial_\mu f_2 -
\mu \frac{f_2}{r} - \\ \nonumber &&
- \partial_r f_3 - \frac{1}{r}\partial_\mu f_4 \Bigr) \\[0.2cm] \nonumber
\frac{1}{2} \frac{\partial D}{\partial t} &=&
- \Bigl( \frac{1}{r} \partial_\mu f_1 + \partial_r f_2 + \frac{f_2}{r} \Bigr) .
\end{eqnarray}
We restrict ourselves to search for stationary solutions. In this case, the time derivatives are equal to zero, and the system
(\ref{E:evolution}) can be solved exactly.
Its solution depends on one arbitrary function $g(r,\mu)$
\begin{eqnarray}\label{E:sol_f_i}
f_1 &=& -g'_r \\[0.2cm] \nonumber
f_2 &=& \frac{1}{r}g'_\mu \\[0.2cm] \nonumber
f_3 &=& g'_r - \frac{\mu}{r}g'_\mu - \frac{g}{r}  \\[0.2cm] \nonumber
f_4 &=& \mu g'_r + \frac{1-\mu^2}{r}g'_\mu - \frac{\mu g}{r}.
\end{eqnarray}
From Eq.~(\ref{E:def_f_i}) and (\ref{E:sol_f_i}), taking into account two relations
(\ref{E:connections}), we get six first-order equations for five unknown functions
$A, B, C, \tilde D, g$.
The single boundary condition is $A, B, C, D \to 0$ when $r \to \infty$ because the pair correlation function of magnetic field fluctuations
should vanish at large scales.

The system of six equations for five unknown functions seems overdetermined and has no solutions at first glance.
However, if $H \gg 1$ this system can be simplified considerably and allows the analytical solution, which is given in
section \ref{SS:sol_big_H}. In this limit case the system is degenerate and is not overdetermined. This suggests that the same would
takes place in the general case.

\subsection{Small mean field $H$}\label{SS:deriv_iso_H}
Because of the large complexity of equations in the anisotropic case, we consider firstly a simpler problem, we assume that the
correlator $\langle b_i b_j \rangle$ is isotropic, i.e. it has the form (\ref{CorrSymB}) even in the case $H \ne 0$.
This can be done in the case of small mean field $H$, because, as we will see below, if $H^2 \ll \tau_c \rho_i \mu_{in} v_0^2 $,
the amplitude of the fluctuations $b_0 \sim H^{1/2}$ is greater than $H$.

To get isotropic equations we have to replace
\[ h_i h_j \Longrightarrow \frac{1}{3} \delta_{ij}. \]
Evolution equation becomes
\begin{equation}\label{E:iso_H_r}
\begin{split}
\frac{1}{2\tau_c} \frac{\partial Q(r)}{\partial t} =&
\left(V(0) - V(r) + \lambda + \frac{2 a}{\tau_c} \frac{H^2}{3}\right) (Q'' + \frac{4Q'}{r}) - \\
& - V' Q' - \frac{1}{r}(4V' + r V'') (Q+\frac{1}{6}H^2 ) ,
\end{split}
\end{equation}
where as before
\begin{equation}\label{E:def_lambda}
\lambda  =\frac{2 a}{\tau_c} (2 Q(0)+\frac{2}{3} r Q'(r)|_{r \to 0}).
\end{equation}
When $H=0$ Eq.~(\ref{E:iso_H_r}) turns into Eq.~(\ref{E:iso_r}).

\section{The solution of equations}
\subsection{Preliminaries}
Let us reduce our equation to dimensionless one. Before that we denote the rms velocity of the neutral gas by $v_0$,
$v_0^2~=~\langle v_i({\w x}) v_i({\w x}) \rangle = 6 V(0) $, and the rms amplitude of the magnetic field fluctuations by $b_0$,
$b_0^2 = \langle b_i({\w x})b_i({\w x}) \rangle$.
The correlation time of the velocity field assumed to be equal to
$\tau_c = \tau_{max} = L_0 / v_0$. That is the eddy turnover time at scale $L_0$.
For length unit we take the size of the molecular cloud $L_{0}$, for unit of time we take $\tau_{max}$, and for unit of magnetic field
we take the value of \[ B_{unit} = \left(\frac{\tau_c}{2 a}V(0) \right)^{1/2} = \left( \frac{\pi}{3} L_0 v_0 \rho_i \mu_{in} \right)^{1/2}. \]
In other words, we introduce new functions and variables
\begin{equation}\label{E:obezrazmerivanie}
\begin{split}
& r^{(1)} = r/L_0; \quad
t^{(1)} = t / \tau_{max} ; \\
& V^{(1)}(r) = V(r)/V(0); \quad
Q^{(1)}(r) = Q(r)/\left(\frac{\tau_c}{2 a}V(0)\right).
\end{split}
\end{equation}
Thus, we reduce Eq.~(\ref{E:iso_H_r}) to the form (below we omit superscript~(1))
\begin{equation}\label{E:iso_H_r_obezr}
\begin{split}
3 \, \frac{\partial Q(r)}{\partial t} =&
\left(V(0) - V(r) + \lambda + \frac{1}{3} H^2 \right) (Q'' + \frac{4Q'}{r}) - \\
&- V' Q' - \frac{1}{r}(4V' + r V'') (Q+\frac{1}{6}H^2 ) .
\end{split}
\end{equation}
where $\lambda  =2 Q(0)+\frac{2}{3} r Q'(r)|_{r \to 0}$, and $V(0)=1$.
For typical parameters of molecular clouds
\begin{equation}\label{paramMC}
\begin{split}
&N_n \simeq  10^{3} \,\text{cm}^{-3} ; \quad
N_i \simeq 10^{-2} \, \text{cm}^{-3}; \quad \\
&L_{0} \simeq 3\cdot 10^{18} \, \text{cm}; \quad \,
v_0 \simeq 1 \, \text{km/s}; \quad T \simeq 50 K,
\end{split}
\end{equation}
the unit of magnetic field is equal to $B_{unit}~\simeq~200 \, \mu G$. But the cloud parameters change in a wide range,
so values of $ B_{unit}$ for them can vary significantly.

Let us estimate the characteristic scales of our problem. The
inertial range of the turbulence is (in the dimensionless
variables)
\begin{equation}
l_\nu < r < 1,
\end{equation}
where $l_\nu = Re^{-3/4} =10^{-6}$ is the viscous scale. Since in the stationary case the magnetic
field energy on one hand is less than the kinetic energy of a neutral gas, but on the other hand is
larger than the kinetic energy of an ionized gas, then
\begin{equation*}
10^{-6} \simeq \frac{\omega_{min}}{\mu_{in}} = (\tau_c \mu_{in})^{-1} \ll \lambda <
(\tau_c \mu_{ni})^{-1} = \frac{\omega_{min}}{\mu_{ni}} \simeq 10.
\end{equation*}
Therefore, we can consider $l_\nu \ll \lambda^{1/\alpha}$.

Let us examine the dimensionless equation (\ref{E:iso_H_r_obezr}).
The parameter $\lambda$ contains the value of $Q(0)$, so
this equation is nonlinear. Let us suppose that the initially magnetic
field fluctuations are weak (and the parameter $\lambda$ is small).
Then the term, which is proportional to $H^2$, makes the positive
contribution to the value of $\partial Q/\partial t$, and lead 
to the initial growth of the magnetic field. Further, this
growth will be stopped by nonlinear terms. So we restrict ourselves
looking for stationary solutions only. Investigation of a
stability of these solutions, especially for anisotropic 
equations (\ref{E:def_f_i}), (\ref{E:evolution}),
is beyond the scope of our paper.

Since for the stationary solution the parameter $\lambda$ does not depend
on time, we will initially consider it to be a constant, which is not connected with the
function $Q(r)$. The single restriction is $\lambda>0$, because of
$3 \lambda = b_0^2$ (in dimensionless variables). Under such
approach the equation (\ref{E:iso_H_r_obezr}) becomes linear.

The similar linear equation arises in the problem of magnetic
dynamo in a turbulent conducting media. In this problem the
parameter $\lambda$ corresponds to the magnetic viscosity and is
considered to be known. The equation, coinciding with Eq.~(\ref{E:iso_H_r_obezr}), 
was investigated in large number of papers
beginning from the paper by Kazantsev~(1968), where the integral equation, 
similar to Eq.~(\ref{E:iso_k}), was derived. The
brief literature review and necessary references are given in the
Discussion. In most of the papers the mean magnetic field was considered
to be zero. In order to reveal the role of the mean magnetic field we
discuss separately cases $H=0$ and $H \ne 0$.

To resolve our equations we need to determine the function $V(r)$ which characterizes
the motion of a neutral gas. The value of
$L_{0}$ corresponds to the maximum scale of the turbulence. It means that the correlation of gas
velocities vanishes at the such scale. We consider the velocity spectrum of a neutral gas to be the
power law. So, the correlation function of gas velocities is
\begin{equation}\label{E:choice_V}
V(r) = \begin{cases}
1 - r^\alpha, & r<1  \\
0,       & r > 1 .
\end{cases}
\end{equation}
Hereinafter we consider the Kolmogorov turbulence.
In this case velocity fluctuation on the scale $r$ is $v(r) \sim
r^{1/3}$ in the inertial range, hence $\alpha=2/3$. On the viscous scales $ r <
l_\nu$ velocity fluctuations are $v(r) \sim r$. Therefore the
correlation function is $V(r)=~1-Cr^2$.  Because $l_\nu =
10^{-6}$ and we consider $(l_\nu)^\alpha \ll \lambda$, then in
this range the correlation function is approximately constant, $V(r) \approx 1$, and
the sought correlation function of the magnetic field is, $Q(r)
\approx Q(0)$, too. The exclusion is solution $Q(r)$ which have
singularity at $r=0$. But we do not consider such solutions (see
below). Thus, the viscous range of scales does not affect the
further analysis.

\begin{figure}
\includegraphics[width = 8cm]{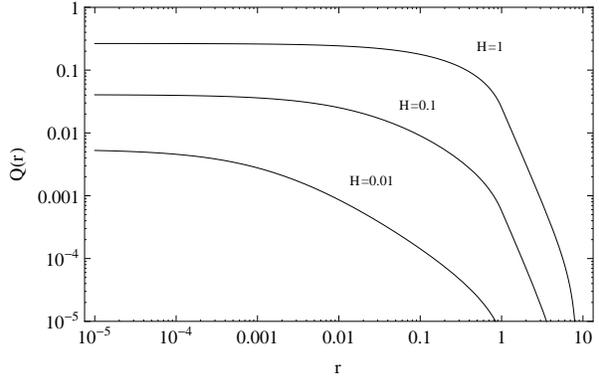}
\caption{The preferential solution $Q(r)$ for different values of the mean magnetic field $H$}\label{F:Q}
\end{figure}

Let us note that in the Kazantsev-Kraichnan model the turbulent velocity is assumed to be $\delta$-correlated in time 
(\ref{E:delta_corr}) with some value of the correlation time $\tau_c$. Because $\tau_c$ characterizes the total realization of the turbulent motion, it formally can not be a function of the scale $r$. 
However, many authors who apply the Kazantsev-Kraichnan model to the problem of magnetic dynamo, in order to approach the physical reality, consider $\tau_c$ to be a function of the scale. They assume $\tau_c$ to be equial to the turnover time at given scale, 
$\tau_c (r) \sim r/ \langle v^2 \rangle^{1/2}_r$. 
In this case $V(r) \sim r \langle v^2 \rangle^{1/2}_r$
and the Kolmogorov turbulence corresponds to the value $\alpha=4/3$.
This assumption is justified by the comparison of theoretical results in the model $\alpha=4/3$
with numerical simulations of forced Navier–Stokes equation (Mason et al. 2011; Tobias,Cattaneo \& Boldyrev 2013). 

From the theoretical point of view, Vainshtein \& Kichatinov (1986) and Boldyrev \& Cattaneo (2004) state that we need to know only
the integral of the velocity correlation function over time, that is, the turbulent diffusivity, which in Kolmogorov
turbulence scales as $r^{4/3}$. 

In current paper we use the Kazantsev-Kraichnan model~(\ref{E:delta_corr}) considering $\tau_c = \text{const}(r)$. 

\begin{figure}
\includegraphics[width=8cm]{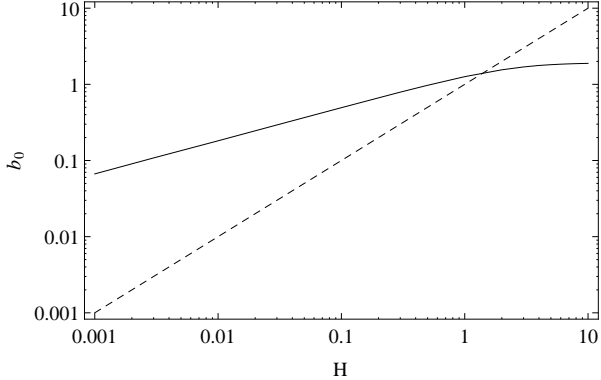}
\caption{The dependence of $b_0$ on $H$. The dashed line $b_0=H$ is given for orientation}\label{F:b0_H}
\end{figure}

\subsection{The case with zero mean field}\label{SS:sol_iso}

To begin with we consider the case of zero mean magnetic field. We put  
$\partial Q/\partial t =0$ and $H=0$ in Eq.~(\ref{E:iso_H_r_obezr}) and we obtain
second order equation for the function $Q(r)$
\begin{equation}\label{E:ODE}
\left(V(0) - V(r) + \lambda \right) (Q'' + \frac{4Q'}{r}) - V' Q' - \frac{1}{r}(4V' + r V'')Q =0,
\end{equation}
where $\lambda=2 Q(0)+\frac{2}{3} r Q'(r)|_{r \to 0}$. In this expression the term
$rQ'(r)$ is essential only for solutions which have singularity at $r=0$. The boundary condition 
is $Q \to 0$ at $r \to \infty$.

Eq. (\ref{E:ODE}) has two independent solutions. At $r \to 0$
($l_\nu \ll r \ll \lambda^{1/\alpha}$) one can find the asymptotics of the
solutions, and at $r>1$ Eq.~(\ref{E:ODE}) can be solved exactly
\begin{eqnarray}\label{E:asimptotics}
r \to 0 &:& \qquad Q_1 \sim 1 ; \qquad Q_2 \sim r^{-3} \\ \nonumber
r >1 &:& \qquad Q_1 = 1; \qquad Q_2 = r^{-3}.
\end{eqnarray}
Now let us discuss the existence of the solution which is bounded at
$r=0$ and is decreasing at $r \to \infty$. At the same time we can
consider more general nonstationary equation, where we assume $\lambda$ 
to be independent of time constant as before. Making the substitution
\begin{equation}
Q(r,t) = \frac{\Psi(r)}{r^2} (m(r))^{1/2} \,e^{-E t},
\end{equation}
where $m(r)^{-1} = V(0) - V(r) + \lambda >0$, we get for the function $\Psi(r)$ the
Shr\"{o}dinger equation with the variable mass $m(r)$
\begin{equation}
\frac{1}{m(r)}\frac{d^2 \Psi}{dr^2} + (E-U(r))\Psi =0.
\end{equation}
The such substitution was firstly done by Kazantsev (1968). The
reduction to the Shr\"{o}dinger equation was discussed in detail
in the paper by Schekochihin, Boldyrev \& Kulsrud (2002). The stationary solution
corresponds to $E=0$, the negative values of the energy $E$ mean
the exponential growth of the magnetic field. For the chosen
velocity correlator $V(r)$ (\ref{E:choice_V}) one can obtain the
analytic expression for the potential $U(r)$. For $\alpha <0.915$
(that includes the Kolmogorov turbulence) the potential $U(r)$ is
positive everywhere. Hence there is no solutions exponentially growing with time.
Eq.~(\ref{E:ODE}) also has no finite at $r=0$ and
vanishing at $r\to\infty$ solution. Consequently 
any solution $Q(r)$ has the same power asymptotics (\ref{E:asimptotics}) 
at $r\to 0$ and at $r\to \infty$. The
solution $Q_1$ of the stationary equation, which is finite at
$r=0$, falls down at $r \to \infty$ not to zero, but to some
positive value. We demand $Q \to 0$ at $r \to \infty$. 
Only the solution $Q_2$, which has singularity at $r=0$, satisfies this condition. This
solution has the power law behavior $r^{-3}$ even at
scales $r < l_\nu$, up to very small
scales where the magnetic viscosity becomes
important. Thus, there is no finite solutions with the zero mean
magnetic field for the Kolmogorov turbulence ($\alpha = 2/3$).

For $\alpha >0.915$  and $\lambda < \lambda_{max}(\alpha)$ there
exists the region where $U(r)<0$. There appears the bound states, i.e.
finite solutions of the Shr\"{o}dinger equation with $E<0$. That
means that the magnetic field will grow up to the level when
$\lambda = \lambda_{max}$.

\begin{figure}
\includegraphics[width=8cm]{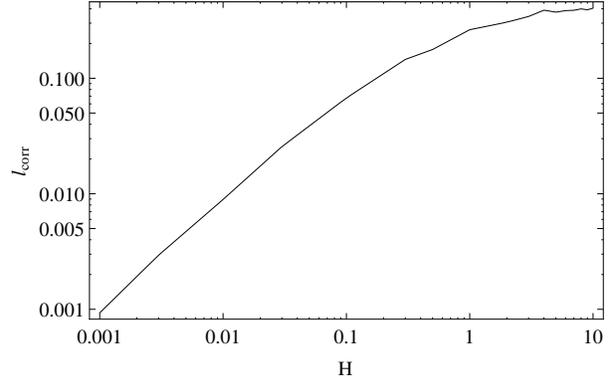}
\caption{The dependence of the correlation length $l_{corr}$ on $H$}\label{F:L_corr}
\end{figure}

\subsection{Small mean field $H$}\label{SS:sol_iso_H}

Now we look for stationary solutions of Eq.~(\ref{E:iso_H_r_obezr}) with small but nonzero mean magnetic field.
This equation is inhomogeneous one, and its partial 
solution is a constant, $Q(r) = -H^2/6$. We introduce the quantity
\begin{equation}
\tilde Q  = Q + \frac{1}{6}H^2,
\end{equation}
denote $\lambda' = \lambda + \frac{1}{3}H^2$, and obtain for the $\tilde Q$ 
exactly the homogeneous equation (\ref{E:ODE}) with the replacement $\lambda \rightarrow \lambda'$.
In fact, the quantity $\tilde Q$ is the correlation function of the total magnetic field $\w B=H+b$
\begin{equation}
\langle B_i({\w x})B_j({\w x+r}) \rangle =
\langle b_i({\w x})b_j({\w x+r}) \rangle + \frac{1}{3} H^2 \delta_{ij}.
\end{equation}
The last term is written in the form $\langle H_i
H_j \rangle = H^2 \delta_{ij}/3$, because in this subsection we assume
correlators to be isotropic. Therefore, the tensor structure of the
correlator $\langle B_i B_j \rangle$ coincides with
(\ref{CorrSymB}). It is possible to derive evolution equation directly
for this correlator, including the mean magnetic
field, as was done by Boldyrev, Cattaneo \& Rosner (2005).
However, the boundary conditions for $\tilde Q$ are different
\begin{eqnarray}
&& 2 \tilde Q(0) +\frac{2}{3} r \tilde{Q}'|_{r \to 0} = \lambda' \\
&& \tilde Q \to \frac{1}{6}H^2, \qquad r \to \infty, \label{E:bound_cond}
\end{eqnarray}
because of the fluctuations correlator tends to zero at large
distances, but the correlator $\langle H_i H_j \rangle$ is
independent on the distance due to its homogeneity.

\begin{figure}
\includegraphics[width=8cm]{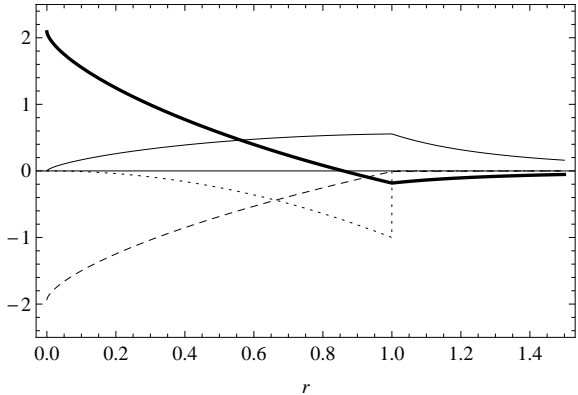}
\caption{Functions $A_0(r)$-thick line, $B_0(r)$-thin, $D_0(r)$-dashed, $D_1(r)$- dotted lines}\label{F:1}
\end{figure}

As was mentioned in the previous subsection, the solution of the
stationary equation (\ref{E:ODE}), which is finite at $r=0$, tends
to a positive constant at $r\to\infty$. It is in accordance with
the new boundary condition (\ref{E:bound_cond}). Now even for
$\alpha=2/3$ there exists the unique bounded solution, we will
call it as the preferential solution. The value of 
$\lambda$ for them is defined by unique manner. Graphs of this solution for
three values of the mean magnetic field $H=1$, $H=0.1$ and $H=0.01$ are
presented on the Fig.~\ref{F:Q}. These graphs and all other numerical 
results are obtained for the Kolmogorov turbulence,
$\alpha=2/3$. Since
\begin{equation}
\langle b_i({\w x})b_i({\w x}) \rangle = b_0^2 = 3 \lambda,
\end{equation}
we get the dependence of $b_0$ on $H$, which is presented on the
Fig.~\ref{F:b0_H}. 
For $H < 1$ the amplitude of the
fluctuating magnetic field $b_0 \sim \sqrt{H}$. Thus, for $H \ll 1$
we have $b_0 \gg H$, i.e. the fluctuating field dominates over the
mean field.

Let us note that for $H\to 0$ the amplitude of the magnetic fluctuations
$b_0$ for the preferential solution tends to zero also. For $H=0$
the preferential solution turns to the zero solution of equation without the
mean field. It corresponds to the fact that for $H=0$ there are no
bounded solutions vanishing at large scales.

For small values of $r$, $r \ll 1$, the asymptotic behavior of the preferential solution is 
\mbox{$Q(r) \simeq Q(0) ( 1- r^\alpha/ \lambda)$}. 
In the k-space it gives $k^2 \overline Q(k)
\sim k^{-5/3}$ for $\alpha=2/3$, i.e. the magnetic fluctuations spectrum
coincides with the spectrum of the hydrodynamic turbulence
of a neutral gas.

For the preferential solution one can also calculate the correlation length of the fluctuating magnetic field,
which we define by the formula
\begin{equation}\label{defLcorr}
l_{corr} = \frac{1}{b_0^2} \int_0^\infty \langle b_i({\w 0})b_i({\w r}) \rangle dr.
\end{equation}
The numerical results are shown on the Fig.~\ref{F:L_corr}. One may
note that $l_{corr} \sim H$ for small values of $H$.

\begin{figure}
\includegraphics[width=8cm]{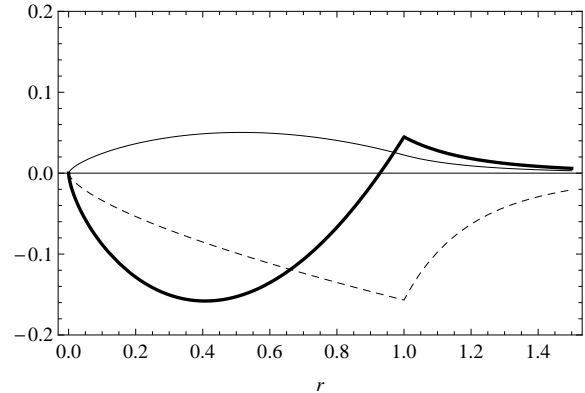}
\caption{Functions $C_0(r)$-thick line, $A_1(r)$-thin, $B_1(r)$-dashed lines}\label{F:2}
\end{figure}

\subsection{Large mean field $H$}\label{SS:sol_big_H}
\begin{figure}
\includegraphics[width=8cm]{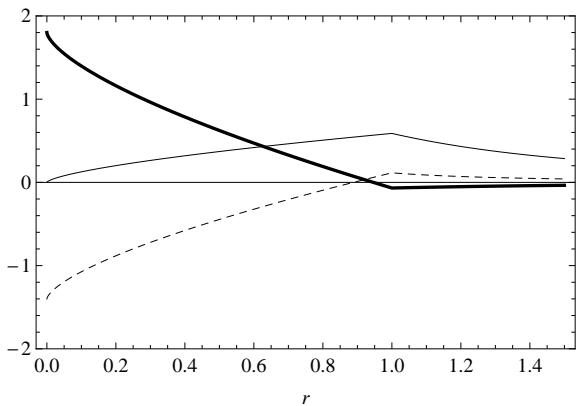}
\caption{Functions $A_0(r)$-thick line, $B_0(r)$-thin, $D_0(r)$-dashed lines
for the smooth correlator $V(r)$}\label{F:3}
\end{figure}

\begin{figure}
\includegraphics[width=8cm]{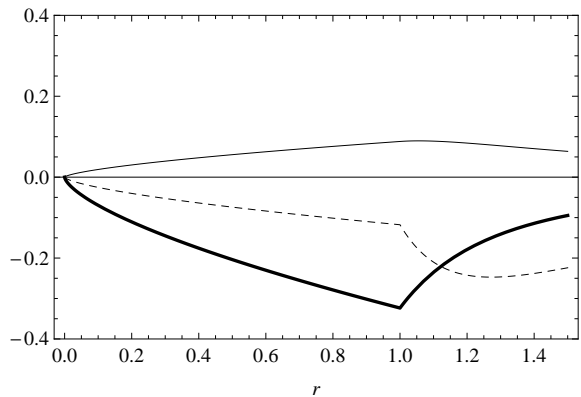}
\caption{Functions $C_0(r)$-thick line, $A_1(r)$-thin, $B_1(r)$-dashed lines for the smooth
correlator $V(r)$}\label{F:4}
\end{figure}

For large mean field, $H \gg 1$ (in dimensionless units), the
correlators of the fluctuating magnetic field can not be considered as
isotropic. We need to solve the system of anisotropic equations
(\ref{E:connections}),(\ref{E:def_f_i}) and (\ref{E:sol_f_i}). To reduce this system to dimensionless 
variables (\ref{E:obezrazmerivanie}) one must change only the
relations (\ref{E:def_XY}) for values $X, Y$:
\begin{eqnarray*}
X &=& D(0)+H^2 =  \tilde D(0) \\
Y(r) &=& -2 A(0)+ 2 V(r) - 2 V(0) +rV'.
\end{eqnarray*}
We assume the correlators $A, B, C, D$ to be the order of unity, and keep only terms of
$H^2 \gg 1$ order. The system of equations is simplified significantly. The solution is described
in detail in the Appendix \ref{SS:append}. It turns out that this system of equations is not overdetermined.
The dependence of the correlation functions over the angle $\theta$ is simple
\begin{eqnarray}
A(r, \mu) &=& A_0(r)+A_1(r)\mu^2 \\ \nonumber
B(r, \mu) &=& B_0(r)+B_1(r)\mu^2 \\ \nonumber
C(r, \mu) &=& C_0(r) \mu \\ \nonumber
D(r, \mu) &=& D_0(r)+D_1(r) \mu^2.
\end{eqnarray}
where  $\mu=\cos \theta$. If we take the correlator $V(r)$ in the form
(\ref{E:choice_V}), as before, then we obtain a solution which is presented on
Fig.~\ref{F:1},~\ref{F:2}. We see that the function $D_1(r)$ is
discontinuous at $r=1$. It is due to the break of the
function $V(r)$ at $r=1$, (\ref{E:choice_V}). If we choose the
correlator $V(r)$, which has the continuous derivative at $r=1$,
then values of $A, B, C, D$ will be also continuous, and the
function $D_1(r)=0$ (see the Appendix \ref{SS:append}). The obtained solution in this case
is shown on the Fig.~\ref{F:3},~\ref{F:4}. For small values of $r \ll 1$
the solution include terms $r^\alpha$ and $r^2$. In
the $k$-space they correspond to terms $k^{-11/3}$ and $k^{-5}$ in $\overline Q(k)$.

\section{Discussion}
The problem of the kinematic dynamo in a conducting medium (hereinafter - DCM) 
is discussed widely. See, for example, papers: Schekochihin et
al. (2002);  Boldyrev \& Cattaneo (2004); Rogachevskii \& Kleeorin (1997); 
Kleeorin \& Rogachevskii (2012), Schleicher et al. (2013) and references therein. 
Also the problem of magnetic field generation in weakly ionized incompressible
gas was studied by Subramanian (1997). 
For analytical treatment the
model of Kazantsev-Kraichnan (Kazantsev 1968; Kraichnan 1968) is
usually used. At the initial stage of the
magnetic field growth its influence on the medium motion is
negligible. Therefore, our problem and the problem of DCM are similar
in many aspects. In the problem of DCM
the evolution equation for the magnetic field is
\begin{equation}\label{E:dBdt_conduct}
\frac{\partial \w B}{\partial t} = {\w \nabla \times(v \times B)} + \nu_m \Delta \w B,
\end{equation}
where $\nu_m$ is the magnetic viscosity. Our Eq.~(\ref{E:dBdt})
contains the additional term, which is proportional to $B^3$,
due to the friction between neutral and ionized gases. The term
$\nu_m \Delta \w B$ in our case is much smaller
than the term $B^3$ because of high conductivity, and we omit it.
Thus, Eq.~(\ref{E:dBdt_conduct}) and Eq.~(\ref{E:dBdt})
differ by the last terms only.  So, evolution equations for the
magnetic field correlators used in papers cited above
almost identical with our Eq.~(\ref{E:iso_r}). Instead of unknown parameter
$\lambda$ (\ref{E:def_lambda}) (it is the averaged square of the
amplitude of magnetic field fluctuations) equations of DCM
contain the magnetic viscosity $\nu_m$ which is considered to be known.
Equations of DCM theory are linear, and the problem is
to find solutions exponentially growing in time. The dependence of
the largest growth rate on $\nu_m$ is discussed in
Kleeorin \& Rogachevskii (2012), Schleicher et al. (2013).
Our equation (\ref{E:dBdt}) and evolution equation for
correlators (\ref{E:iso_r}) are nonlinear because the value of
$\lambda$ is defined by the solution. This nonlinearity stops
the initial growth of magnetic fluctuations. Therefore we are looking for
stationary solutions only.

The main result of our paper is taking into account the mean
magnetic field, and derivation of evolution equations for
anisotropic correlators $\langle b b\rangle$. In all mentioned
papers magnetic field correlators were suggested to be isotropic, as in
section \ref{SS:deriv_iso_H} of current work. Rogachevskii \& Kleeorin (1997)
tried to take into account
a mean magnetic field. But when they derive the evolution equation for
pair correlators, they put mean field to zero.
Boldyrev et al. (2005) made no
assumption of zero mean magnetic field. They considered the pair
correlator of the total magnetic field $\tilde Q \simeq \langle B
B\rangle $. Obtained evolution equation for $\tilde Q$ coincides
with Eq.~(\ref{E:iso_r}) of our paper for $Q$ (see section
\ref{SS:sol_iso_H}). However, the boundary conditions for $\tilde
Q$ are different. We consider the media is uniform and the
mean magnetic field is constant, so the correlator 
$\langle H({\w 0}) H({\w r}) \rangle$ is independent of ${\w r}$.
At the same time, the correlator of magnetic field fluctuations
must vanish at large scales. So, $Q(r) \to 0$, but
 $\tilde Q(r) \to H^2/6$ at $r \to \infty$. As we show in the section 
 \ref{SS:sol_iso_H}, this difference is essential for existence of
bounded solutions.

In all mentioned papers the velocity correlation function was
assumed to be $V(r) = 1-r^\alpha$. In the model with scale-independent correlation time $\tau_c$ 
for the Kolmogorov turbulence $\alpha=2/3$, and, as we show, there are no solutions
exponentially growing in time, see also Rogachevskii \& Kleeorin (1997). 
If $\tau_c$ is supposed to be a function of the scale, the Kolmogorov
turbulence corresponds to $\alpha=4/3$ and growing solutions of
(\ref{E:iso_r}) appear. This is equivalent to the appearance of
solutions of stationary equation (\ref{E:ODE}) bounded at $r=0$
and vanishing at large scales. In current work we demonstrate that when
mean field is taken into account, bounded solutions do exist even in the model
with $\alpha =2/3$.

\section{Conclusion}
We find the correlation function of fluctuating magnetic field
when the mean magnetic field equals zero, $H=0$. For the
Kolmogorov turbulence 
the correlation function diverges as $r^{-3}$ when $r\to 0$ up to small resistive scales.

For the case $H \ne 0$ the problem is solved in the approximation
of isotropic correlators. It is shown that for each value of $H$
there exists the unique solution bounded at $r=0$. When $H<1$ the amplitude
of the fluctuating field $b_0$ turns out to be proportional to the square root
of the mean field, $b_0 \simeq H^{1/2}$. Therefore, for $H \ll 1$
correlators can be considered to be isotropic. However, this solution
exists only at the presence of the finite, even weak, mean
magnetic field $H$. At $L \ll L_0$ the magnetic field correlator
has the same form as that of the turbulent neutral gas velocity,
$Q(r) = Q(0) ( 1 - r^\alpha/ \lambda)$. In the $k$-space it
corresponds to the spectrum $\overline Q(k) \sim k^{-11/3}$.

Also we derive exact anisotropic equations for the magnetic field
correlators at the presence of the mean magnetic field. We find
the analytic solution at the limit of large mean field $H \gg 1$.
The anisotropic correlation function of the magnetic field
contains the dependence on the angle $\theta$ between the mean
magnetic field and the vector ${\w r}$ in the form of 
$\cos \theta $, and $\cos^2 \theta$ only. The dependence of the
correlators on $r$ at $r<1$ is the sum of two power law
functions, $r^\alpha$ and $r^2$. 

\section*{Acknowledgements}

This work was done under support of the Russian Foundation
for Fundamental Research grant 11-02-01021 12.  A. Kiselev is supported in parts also by the grant 12-02-31648 and the
LPI Educational-Scientific Complex.

\appendix

\section{Large mean field}\label{SS:append}
Here we describe in detail the solution of the system of anisotropic equations
(\ref{E:connections}),(\ref{E:def_f_i}) and (\ref{E:sol_f_i}) in the case of $H \gg 1$.

Let us denote $\tilde g = g/H^2$. We suppose that the correlators $A, B, C, D$ are of the order of
unity and keep only terms of $H^2~\gg~1$ order, and we get the system of equations
\begin{eqnarray}\label{E:syst_H_inf}
\nonumber
&& r A'_r - \mu A'_\mu + r B'_r +r \mu C'_r + (1-\mu^2)C'_\mu + 2 B - \mu C =0  \\
\nonumber
&& A'_\mu +r C'_r + r \mu D'_r + (1-\mu^2) D'_\mu + 3C = 0 \\
\nonumber
&& r \mu A'_r + (1-\mu^2)A'_\mu - \mu B-C -r \mu (3V'+r V'') = -r \tilde g'_r\\
\nonumber
&& -r A'_r +\mu A'_\mu + B +C'_\mu - r D'_r + \mu D'_\mu + r V' = \tilde g'_t\\
\nonumber
&& B'_\mu -r C'_r + \mu C'_\mu + C -r \mu (V'-r V'')= r \tilde g'_r - t \tilde g'_\mu - \tilde g \\
&&r \mu \tilde g'_r + (1-\mu^2) \tilde g'_\mu - \mu \tilde g = 0.
\end{eqnarray}
From the last equation and the fact that the function $\tilde g$ must be odd with respect to $\mu$
it follows $\tilde g=0$. So we have five equations on four functions. Let us note that only 
the derivatives of the function $D$ enters into the system (\ref{E:syst_H_inf}), and only
in the second and in the fourth equations. One can express the values of $D'_r$, $D'_\mu$
\begin{eqnarray}\label{derivD_complex}
\nonumber
D'_\mu &=&  \mu r A'_r - (1+\mu^2) A'_\mu - \mu B - r C'_r - \mu C'_\mu - 3 C - \mu r V' \\ \nonumber
r D'_r &=& -(1-\mu^2) r A'_r - \mu^3 A'_\mu + (1-\mu^2) B - \mu r C'_r + \\
&& +(1-\mu^2) C'_\mu - 3 \mu C + (1-\mu^2) r V'.
\end{eqnarray}
Using the first and the third equations of the system (\ref{E:syst_H_inf}), expressions
(\ref{derivD_complex}) can be simplified
\begin{eqnarray}\label{derivD}
\nonumber
D'_\mu &=&  -2 A'_\mu - r C'_r - \mu C'_\mu - 2 C +\mu r (2V'+r V'') \\ \nonumber
r D'_r &=& -2 r A'_r - r B'_r - B - 2\mu r C'_r - \mu C + r V' \\
&&+ \mu^2 r (2V'+r V'').
\end{eqnarray}
Using equations (\ref{E:syst_H_inf}), one
can show that mixed derivatives of $D$ calculated from
(\ref{derivD}), are the same
$\partial/\partial \mu D'_r = \partial/\partial r D'_{\mu}$. So the function $D$ is correctly
defined by Eq.~(\ref{derivD}). Thus, we have three equations for
three functions $A,B,C$,
\begin{eqnarray}\label{E:last_three}
\nonumber
&& r A'_r - \mu A'_\mu + r B'_r +r \mu C'_r + (1-\mu^2)C'_\mu + 2 B - \mu C= 0\\ \nonumber
&& r \mu A'_r + (1-\mu^2)A'_\mu - \mu B-C  = r \mu (3V'+r V'') \\
&& B'_\mu -r C'_r + \mu C'_\mu + C = r \mu (V'-r V'').
\end{eqnarray}
One can solve this system and after that calculate the function $D$ using (\ref{derivD}). So the system
(\ref{E:syst_H_inf}) is not overdetermined.

Since functions $A, B$ are even with respect to $\mu$, but $C$ is odd, we are looking for the solution in the form
\begin{eqnarray}\label{E:parabola}
A(r, \mu) &=& A_0(r)+A_1(r)\mu^2 \\ \nonumber
B(r, \mu) &=& B_0(r)+B_1(r)\mu^2 \\ \nonumber
C(r, \mu) &=& C_0(r) \mu.
\end{eqnarray}
Substituting these relations into Eq.~(\ref{E:last_three}) we get the system of five ordinary differential
equations for five functions $A_0, B_0, C_0, A_1, B_1$
\begin{equation}\label{E:syst_ODE}
\begin{pmatrix}
\hat \beta & \hat \beta + 2 & 1 & 0 & 0 \\
0 & 0 & \hat \beta - 2 & \hat \beta - 2 & \hat \beta + 2 \\
\hat \beta & -1 & -1 & 2 & 0 \\
0 & 0 & 0 & \hat \beta - 2 & -1 \\
0 & 0 & 2 - \hat \beta & 0 & 2
\end{pmatrix}
\begin{pmatrix}
A_0 \\ B_0 \\ C_0 \\ A_1 \\ B_1
\end{pmatrix}
=
\begin{pmatrix}
0 \\ 0 \\ r (3V'+r V'')  \\ 0 \\ r (V'-r V'')
\end{pmatrix}
\end{equation}
Here we introduce the notation $\hat \beta = r \partial_r$.
The solution of such system is the sum of the general solution of the homogeneous system
(with zero right hand side) and the partial solution of inhomogeneous system. We search for solution of the homogeneous system as a power law function, since it is
the eigen function of the operator $\hat \beta$,
\[ \hat \beta r^\beta = \beta r^\beta. \]
Replacing the operator $\hat \beta$ by a number, $\hat \beta \to \beta$,
we obtain the homogeneous system of linear algebraic equations.
To have nonzero solutions this system must be degenerate.
Equating the determinant of the matrix in the left hand side of Eq.~(\ref{E:syst_ODE}) to zero,
we find the eigen values of $\beta$,
\[ \beta_1=0; \quad \beta_2=-3; \quad \beta_{3,4}=2; \quad \beta_5=-5. \]
We find the solution of the system of linear algebraic equations
with these values of $\beta$ and get the general solution of the
homogeneous system from Eq.~(\ref{E:syst_ODE}). If we take the
neutral gas velocity correlator $V(r)$ in the form of a power law
function, we can also find the partial solution of the
inhomogeneous system, and thereby solve the system
(\ref{E:syst_ODE}) analytically. We assume, as before, the
function $V(r)$ to be in the form (\ref{E:choice_V}) and we obtain the
partial solution: zero at $r > 1$ and the following vector at $r <
1$
\begin{equation}\label{sol:Neodnor}
\begin{pmatrix}
A_0 \\ B_0 \\ C_0 \\ A_1 \\ B_1
\end{pmatrix}_{\text{}}
=
 \frac{\alpha}{\alpha+5}
\begin{pmatrix}
-\frac{\alpha^2 +7\alpha+9}{\alpha} \\ \alpha+6 \\
 -(\alpha+3) \\ 1 \\ \alpha-2
\end{pmatrix}
r^{\alpha}  \stackrel{\mathrm{def}}{=}
\vec{F}(\alpha) r^{\alpha}.
\end{equation}
We denote the vector in the right hand side of (\ref{sol:Neodnor}) as $\vec{F}(\alpha)$.

We are looking for the solution bounded at $r=0$ and vanishing at $r=\infty$. Therefore, at
$r < 1$ we keep only two powers $r^0$ and $r^2$; at $r>1$ we keep powers $r^{-3}$ and $r^{-5}$.
The solution, which is continuous at $r = 1$, is
\begin{eqnarray*}
&& r < 1: \\
&& \begin{cases}
A_0 &= F_1(\alpha) r^\alpha  - (3c_3-4c_4) r^2  - c_1 \\
B_0 &= F_2(\alpha) r^\alpha  + (4c_3-2c_4) r^2 \\
C_0 &= F_3(\alpha) r^\alpha  - 10 c_3 r^2 \\
A_1 &= F_4(\alpha) r^\alpha  - 5 c_4 r^2 \\
B_1 &= F_5(\alpha) r^\alpha
\end{cases} \\[0.5cm]
&& r > 1: \\
&& \begin{cases}
A_0 &= 1/5 c_5 r^{-5} + c_2 r^{-3} \\
B_0 &= -c_5 r^{-5} - 3 c_2 r^{-3} \\
C_0 &=  -2c_5 r^{-5} \\
A_1 &= -c_5 r^{-5} \\
B_1 &= 7c_5 r^{-5}
\end{cases}
\end{eqnarray*}
where
\begin{equation}
\begin{split}
&c_1 = -\frac{13\alpha+54}{30}; \qquad c_2 = -\frac{4}{15}\alpha; \qquad
c_3 = -\frac{\alpha}{14}; \\
&c_4 = \frac{\alpha}{35}; \qquad
c_5 = \frac{\alpha(\alpha-2)}{7(\alpha+5)},
\end{split}
\end{equation}
and the vector $\vec{F}(\alpha)$ is defined in (\ref{sol:Neodnor}).

Let us find the function $D(r,\mu)$. Substituting Eq.~(\ref{E:parabola}) into Eq.~(\ref{derivD}),
we get
\[ D(r,\mu) = D_0(r)+D_1(r) \mu^2, \]
where
\begin{eqnarray}
2 D_1 &=& -4 A_1 - (\hat \beta+3)C_0 + r(2 V'+r V'') \\ \nonumber
\hat \beta D_0 &=& -2 \hat \beta A_0 - (\hat \beta+1)B_0 + r V' .
\end{eqnarray}
The function $D_0(r)$ is defined with accuracy up to a
constant. We choose this
constant so that $D_0 (\infty)=0$. The answer is
\begin{eqnarray*}
&& r < 1: \\
&& \begin{cases}
D_0 &= \frac{\alpha^2 +6\alpha+7}{\alpha+5} r^\alpha - 5c_4r^2 - \frac{1}{5} (4 \alpha+7) \\
D_1 &= -\frac{3}{2} \alpha r^2
\end{cases} \\[0.5cm]
&& r > 1: \\
&& \begin{cases}
D_0 &= 2/5 c_5 r^{-5} \\
D_1 &= 0
\end{cases}
\end{eqnarray*}
The graphs of the obtained solution for $\alpha=2/3$ are shown on
Fig.~\ref{F:1},~\ref{F:2}. We see that the function $D_1$ is
discontinuous at $r=1$. It is caused by the break of the
function $V(r)$ at $r=1$. To obtain a continuous
solution we take the velocity correlator in the form
\begin{equation}
V(r) = \begin{cases}
1 - V_0 r^\alpha, &  r < 1\\
-V_1 r^\gamma, &  r > 1,
\end{cases}
\end{equation}
where the power $\gamma<0$ is arbitrary, and constants are
\begin{equation}
V_0 = - \frac{\gamma}{\alpha - \gamma},  \qquad V_1 = - \frac{\alpha}{\alpha - \gamma}
\end{equation}
These constants are chosen for the functions $V(r)$ and
$V'(r)$ to be continuous at $r=1$.
We repeat the procedure described above, and get the solution
\begin{eqnarray*}
&& r < 1: \\
&& \begin{cases}
A_0 &= V_0 F_1(\alpha) r^\alpha - c_1 \\
B_0 &= V_0 F_2(\alpha) r^\alpha \\
C_0 &= V_0 F_3(\alpha) r^\alpha \\
A_1 &= V_0 F_4(\alpha) r^\alpha \\
B_1 &= V_0 F_5(\alpha) r^\alpha \\
D_0 &= V_0  F_6(\alpha) r^\alpha - \frac{7}{5}\\
D_1 &=0
\end{cases} \\[0.5cm]
&& r > 1: \\
&& \begin{cases}
A_0 &= V_1 F_1(\gamma) r^\gamma + 1/5 c_5 r^{-5} \\
B_0 &= V_1 F_2(\gamma) r^\gamma -  c_5 r^{-5} \\
C_0 &= V_1 F_3(\gamma) r^\gamma - 2 c_5 r^{-5} \\
A_1 &= V_1 F_4(\gamma) r^\gamma - c_5 r^{-5} \\
B_1 &= V_1 F_5(\gamma) r^\gamma + 7 c_5 r^{-5} \\
D_0 &= V_1 F_6(\gamma) r^\gamma + 2/5 c_5 r^{-5} \\
D_1 &=0
\end{cases}
\end{eqnarray*}
where we added the function $F_6(\alpha)$ to functions
$F_1(\alpha),.. F_5(\alpha)$ defined in (\ref{sol:Neodnor}),
\[ F_6(\alpha) = \frac{\alpha^2 +6\alpha+7}{\alpha+5}, \]
Now the constants are
\begin{equation}
c_1 = -\frac{9}{5}, \qquad
c_5 = -\frac{\alpha \gamma}{(\alpha+5) (\gamma+5)}.
\end{equation}
For any value of $\gamma$ all correlators are continuous, and the function $D_1=0$.
So, we can take any reasonable value of $\gamma$, for example, $\gamma=-2$.
Graphs of such solution for $\alpha=2/3$ and $\gamma=-2$ are shown on Fig.~\ref{F:3},~\ref{F:4}.

\end{document}